\documentclass[aps,prd,preprintnumbers,amsmath,amssymb,superscriptaddress]{revtex4}
\usepackage{graphicx,graphics,amsmath,amssymb}
\usepackage{subfigure}
\begin{document}
\title{Slow-Roll Inflation in the Presence of a Dark Energy Coupling}
\author{Philippe Brax}
\affiliation{Institut de Physique Theorique, CEA, IPhT,\\
CNRS, URA 2306, F-91191 Gif/Yvette Cedex, France}
\author{Carsten van de Bruck}
\author{Lisa M. H. Hall} 
\author{Joel M. Weller}
\affiliation{Department of Applied Mathematics, University of Sheffield\\
Hounsfield Road, Sheffield S3 7RH, United Kingdom}

\date{10 December 2008}

\begin{abstract}
In models of coupled dark energy, in which a dark energy scalar field 
couples to other matter components, it is natural to expect a coupling 
to the inflaton as well. We explore the consequences of such a coupling
in the context of single field slow-roll inflation. Assuming an exponential 
potential for the quintessence field we show that the coupling to the inflaton 
causes the quintessence field to be attracted towards the minimum of the 
effective potential. If the coupling is large enough, the field is 
heavy and is located at the minimum. We show how this affects the 
expansion rate and the slow-roll of the inflaton field, and therefore the 
primordial perturbations generated during inflation. We further show that 
the coupling has an important impact on the processes of reheating and preheating. 
\end{abstract}

\maketitle

\section{Introduction} \label{sec:intro}
The nature and origins of dark energy, the energy component which is responsible for the observed 
accelerated expansion of the universe, remain a mystery. Although the observations can be accounted 
for by the cosmological constant, scalar fields \cite {Wetterich:1987fm,Ratra:1987rm} and modified gravity theories \cite{Carroll:2004de} have also been suggested 
(for reviews and references see e.g. \cite{Copeland:2006wr, Martin:2008qp, Nojiri:2006ri}). One of the major aims of modern cosmology 
is to determine the properties of dark energy, many of the parameters of which (equation of state, matter couplings etc.) can be 
constrained by considering data from supernovae at high redshifts, observations of 
anisotropies in the cosmic microwave background radiation (CMB) and the large scale structures 
(LSS) in the universe. At even higher redshifts and 
in the early universe, constraints from varying constants and big bang nucleosynthesis (BBN) give 
further constraints on the dark energy evolution \cite{Kneller:2002zh, Dent:2008vd}. Some models of dark energy can even be tested in the laboratory \cite{Brax:2007ak,Brax:2007vm,Brax:2007hi,Stojkovic:2007dw,Greenwood:2008qp}. In this paper, we study the impact of couplings of 
a quintessence-like scalar field to the inflaton field, the scalar field responsible for an accelerated 
expansion in the very early universe. In this important epoch, the seeds for the structures we observe 
in the universe were created. It is usually assumed that dark energy is not important during 
inflation. In the case in which dark energy and the inflaton field are not coupled, the vacuum 
expectation value (VEV) of the dark energy field is driven by quantum fluctuations to large field 
values, but otherwise there are no consequences for the inflationary dynamics (\cite{Malquarti:2002bh}; see also \cite{Martin:2004ba}). We show
that this is not necessarily the case if the inflaton field couples to the dark 
energy field. In models such as coupled quintessence (\cite{Wetterich:1994bg,Amendola:1999er}; see \cite{Martin:2008qp} for a recent review) or quintessence models with a growing matter 
component \cite{Amendola:2007yx}, dark energy couples to at least one species, which is  thought to be the decay product of the 
inflaton field. Therefore,  it is natural in these types of models to consider a coupling between dark energy and the inflaton field  as well. We will show in this paper that for large enough couplings (to be specified below) one 
can expect modifications to the predictions of the spectral index, its running and the tensor-to-scalar ratio. 
We find also that  the details of the physics of reheating and preheating are affected by the presence 
of a coupling between dark energy and the inflaton. The paper is organized as follows: in 
Section \ref{sec:2} we describe our model  and study the inflationary epoch and discuss the 
effect of the quintessence field on the primordial perturbations. In Section \ref{sec:3} we discuss the 
consequences for reheating and preheating. We conclude in Section \ref{sec:conclusions}. 

\section{Slow-roll inflation in the presence of coupled dark energy} \label{sec:2}
The theory we consider in this paper is specified by the action 
\begin{eqnarray}
{\cal S} = \int {\rm d}^4 x \sqrt{-g}  \left[\frac{M_{\rm Pl}^2}{2}{\cal R} + {\cal L}_Q + {\cal L}_\phi\right]\nonumber
\end{eqnarray}
with 
\begin{eqnarray}
{\cal L}_Q &=& -\frac{1}{2}g^{\mu\nu}(\partial_\mu Q)(\partial_\nu Q) - V(Q) \nonumber \\
{\cal L}_\phi &=& -\frac{A^2(Q)}{2}g^{\mu\nu}(\partial_\mu \phi)(\partial_\nu \phi) - A(Q)^4 U(\phi). \nonumber 
\end{eqnarray}
Here, ${\cal R}$ is the Ricci scalar, $\phi$ is the inflaton field, $Q$ is another scalar field, possibly playing the role of 
dark energy and $g$ is the determinant of the metric tensor. The appearance of the coupling function $A(Q)$ follows directly from the fact that we focus on a 
scalar tensor theory where matter, i.e. the inflaton field here, couples to the rescaled metric $\tilde g_{\mu\nu}=A^2(Q) g_{\mu\nu}$.
While these equations are valid for any potential $U(\phi)$, we will 
use as an example the standard chaotic inflation potential
\begin{equation}\label{eq:quad}
U(\phi) = \frac{1}{2}m^2 \phi^2.
\end{equation}
For the coupling function  we choose $A(Q) = \exp(\beta Q/M_{\rm Pl})$, (as in, for example, \cite{Wetterich:1994bg,Amendola:1999er,Bean:2008ac}). 
After a field redefinition, the (effective) mass of the inflaton field is nothing but $A(Q) m$, which grows as $Q$ rolls down along the potential
towards large values. Additionally, we choose the potential for the quintessence field to be an 
exponential potential, i.e.
\begin{equation}\label{eq:pot}
V(Q) = M_{\rm Pl}^4 \exp(-\lambda Q/M_{\rm Pl}), 
\end{equation}
with $\lambda$ positive.  

\subsection{Slow-Roll Inflation} \label{sec:21}
Let us now consider the inflationary period. Since $Q$ couples to $\phi$, the inflationary dynamics
will be modified by the coupling. The coupling can potentially ruin inflation and therefore we first consider 
the conditions on the parameters in order to obtain a period of slow--roll inflation. 

The equations of motion for the fields $\phi$ and $Q$ in a homogeneous and isotropic universe are 
\begin{equation}\label{eq:Qeq}
\ddot Q + 3H \dot Q + \frac{\partial V}{\partial Q} = A^2 \beta (\dot\phi^2 - 4A^2 U),
\end{equation}
\begin{equation}\label{eq:phieq}
\ddot \phi + 3H \dot \phi + A^2\frac{\partial U}{\partial \phi} = -2 \beta \dot Q \dot\phi.
\end{equation}
The Friedmann equation reads 
\begin{equation}
H^2 = \frac{1}{3M_{\rm Pl}^2}\left(\frac{A^2}{2}\dot\phi^2 + \frac{1}{2}\dot Q^2 + V + A^4U \right).
\end{equation}
The $Q$-field moves in an effective potential, which is given by the bare potential and a 
part coming from the coupling to $\phi$. Taking $\lambda$ positive, the sign of $\beta$ determines 
whether the effective potential has a minimum or not. For positive $\beta$, a local minimum exists, whereas 
there is not one in the case of a negative $\beta$. If the minimum does not exist, the effective potential 
is of a runaway form and the discussion of the field behaviour will be similar to that in \cite{Malquarti:2002bh}. 
In this paper, we will assume the existence of a minimum, that is, we consider the case of a positive $\beta$. 
Assuming then slow-roll of the scalar field $\phi$, this determines the value of $Q$ at the minimum of the 
effective potential to be 
\begin{equation}\label{eq:Qmin}
Q_{\rm min} = \frac{1}{4\beta + \lambda} \ln \left(\frac{\lambda M_{\rm Pl}^4}{4\beta U(\phi)} \right),
\end{equation}
together with the condition for the minimum 
\begin{equation}\label{minconst}
\lambda V(Q_{\rm min}) = 4 \beta A^4 U(\phi).
\end{equation}
For later use, we also state $A(Q_{\rm min})$ during slow-roll inflation:
\begin{equation}\label{A}
A(Q_{\rm min}) = \left(\frac{\lambda M_{\rm Pl}^4}{4\beta U(\phi)} \right)^{\beta/(4\beta + \lambda)}.
\end{equation}

During inflation, in which we assume that both of the fields are rolling slowly, 
using eqn.(\ref{minconst}) the Friedmann equation takes the form: 
\begin{equation}\label{eq:friedmann}
3H^2 \approx  \frac{1}{M_{Pl}^2}(A^4U+V) \approx \frac{A^4U}{M_{Pl}^2}\left(1+\frac{4\beta}{\lambda}\right).
\end{equation}
We will justify below the assumption that both fields roll slowly. 
Note that the $Q$-field contributes to the expansion rate with an amount depending on $\beta$ and $\lambda$.
The mass of the $Q$--field can be found to be 
\begin{eqnarray}
m^2_{\rm eff} = \frac{\partial^2 V}{\partial^2 Q} - 2 A^2 \beta^2 \dot\phi^2 + 16 A^4 \beta^2 U.
\end{eqnarray}
In the case of a slowly--rolling $\phi$--field, this gives
\begin{eqnarray}
m^2_{\rm eff} &\approx& \frac{\partial^2 V}{\partial^2 Q} + 16 A^4 \beta^2 U \nonumber \\
&\approx& \frac{\partial^2 V}{\partial^2 Q} + \frac{48 \beta^2 H^2}{1+\frac{4\beta}{\lambda}},
\end{eqnarray}
where in the last line we have used the Friedmann equation. For large enough $\beta$, 
the field is rather heavy ($m^2_{\rm eff}> H^2$) and will therefore settle into the 
minimum of the effective potential. 
 To get a rough idea on how small $\beta$ can be, we can solve the 
equation $m^2_{\rm eff}/H^2=1$ and obtain  $\beta_{\rm crit} = \frac{1}{12\lambda}$ 
(we remind the reader that $\beta$ has to be positive for a minimum to exist, since we 
are assuming $\lambda$ positive). 
This equation implies that, even when  $\beta$ is  rather small, the $Q$-field sits at  the minimum of the 
effective potential. We find numerically that $Q$ settles into the minimum even for $\beta$ as small as 0.05 (see section \ref{sec:cons}).

It is possible to obtain a degree of analytical insight into the behaviour of the system during the inflationary 
period by studying the slow--roll regime and deriving the slow--roll parameters. Firstly, we show that the extra friction term in eqn.~(\ref{eq:phieq}) 
is negligible during inflation. Throughout this analysis, we will assume that the value of $\beta$ is such that Q 
settles into the minimum of the effective potential. Thus one can write $\dot{Q} = \dot{Q}_{\rm min}$. 
Using eqn.~(\ref{eq:Qmin}) we find ($'=d/d\phi$)
\begin{equation}
\dot Q_{\rm min} = -\frac{1}{4\beta + \lambda}\frac{U'}{U} \dot\phi = -\frac{2}{\lambda+4\beta} \left(\frac{\dot{\phi}}{\phi} \right),
\end{equation}
where eqn. (\ref{eq:quad}) was used in the last step. Therefore we find
\begin{equation}
|2\beta \dot{Q}| = \frac{4\beta}{\lambda+4\beta}\left(\frac{\dot{\phi}}{\phi} \right) \ll \dot{\phi} \ll H.
\end{equation}
In general, we see that the extra damping term is proportional to $\dot\phi^2$ and we can therefore neglect it:
\begin{equation}
\ddot \phi + 3H\dot\phi + \frac{\partial U}{\partial \phi}A^2 \approx 0.
\end{equation}
This means that the condition for slow-roll differs from the standard case by a factor of $A^2$:
\begin{equation}\label{eq:slowroll1}
\dot{\phi} \approx -\frac{U' A^2}{3H}.
\end{equation}

The first slow-roll parameter is defined by $\epsilon\equiv-\textstyle\frac{\dot{H}}{H^2}$ and in 
terms of the fields can be found to be 
\begin{equation}\label{eq:epsilon}
\epsilon \approx \frac{M_{\rm Pl}^2}{2 A^2} \left(\frac{U'}{U}\right)^2\left(1+\frac{4\beta}{\lambda}\right)^{-2},
\end{equation}
where eqns. (\ref{eq:slowroll1}) and (\ref{eq:friedmann}) were used. We can find the $\eta$ parameter in a similar way. 
Differentiating eqn.~(\ref{eq:slowroll1}) yields
\begin{equation}
\ddot{\phi} \approx -\frac{1}{3}\epsilon U'A^2 - \frac{U''}{3H}\dot\phi A^2 + \frac{2}{3}\frac{U'^2 A^2}{HU}\frac{\beta}{4\beta+\lambda} \dot\phi.
\end{equation}
The requirement that $|\ddot{\phi}|\ll |H\dot{\phi}|$ gives
\begin{equation}
\left|\epsilon - \eta + \frac{4\beta}{\lambda}\epsilon \right| \ll 1. 
\end{equation}
where
\begin{equation}\label{eq:eta}
\eta\equiv \frac{U''}{U}\frac{M_{Pl}^2}{A^2}\left(1+\frac{4\beta}{\lambda}\right)^{-1}.
\end{equation}

\subsection{Cosmological Perturbations}

We now derive the relevant equations for the cosmological perturbations. We show first that the perturbations in 
the $Q$-field are much smaller than the perturbations in the $\phi$-field. This is because the $Q$-field is a heavy 
field (i.e. $m_{\rm eff}>H$) and it is well known that for these fields perturbations are suppressed. 
The fluctuations of a massive scalar field during inflation satisfy an equation composed of oscillatory and 
non-oscillatory parts \cite{Riotto:2002yw}. The expansion of the universe causes the wavelength of the perturbations 
to be streched, so that the former can be neglected as the average contribution of oscillations averages to zero. The amplitude 
of the resulting power spectrum is suppressed by $(H/m_{\rm eff})$ and decreases rapidly for large wavelengths. 
For completeness, however, we will study numerically the perturbations for both fields. The inflaton field $\phi$ is 
light ($m<H$) and its perturbations are not suppressed. 

In the longitudinal gauge and in the absense of anisotropic stress, the scalar perturbations of the FRW metric can be expressed as (see 
e.g. \cite{Mukhanov})
\begin{equation}
ds^2 = -(1+2\Psi)dt^2+a^2(1 - 2\Psi)\delta_{ij}dx^idx^j.
\end{equation} 
The perturbed Einstein equations give 
\begin{eqnarray}
3H(\dot{\Psi}+H\Psi)+\frac{k^2}{a^2}\Psi = -\frac{1}{2M_{\rm Pl}^2}\delta\rho, \\
\dot{\Psi}+H\Psi = -\frac{1}{2M_{\rm Pl}^2}\delta q, \\
\ddot{\Psi}+4H\dot{\Psi}+(3H^2+2\dot{H})\Psi = \frac{1}{2M_{\rm Pl}^2}\delta p,
\end{eqnarray}
where
\begin{eqnarray}
\delta\rho &=& (\dot{Q}(\delta Q)\dot{}-\Psi\dot{Q}^2)+\beta A^2 \dot{\phi}^2 \delta Q + A^2(\dot{\phi}(\delta \phi)\dot{}-\Psi\dot{\phi}^2)+\left(\frac{\partial V}{\partial Q}\delta Q+4\beta A^4 U \delta Q  + A^4 \frac{\partial U}{\partial \phi}\delta\phi\right), \\
\delta q &=& -(\dot{Q}\delta Q+A^2\dot{\phi}\delta\phi), \\
\delta p  &=& (\dot{Q}(\delta Q)\dot{}-\Psi\dot{Q}^2)+\beta A^2 \dot{\phi}^2 \delta Q + A^2(\dot{\phi}(\delta \phi)\dot{}-\Psi\dot{\phi}^2)-\left(\frac{\partial V}{\partial Q}\delta Q+4\beta A^4 U \delta Q  + A^4 \frac{\partial U}{\partial \phi}\delta\phi\right).
\end{eqnarray}
The perturbed field equations are
\begin{eqnarray}
(\delta\phi)\ddot{} &+& (3H+2\beta\dot{Q})(\delta\phi)\dot{}+\left(\frac{k^2}{a^2}+A^2\frac{\partial^2U}{\partial\phi^2}\right)\delta\phi = -2\beta\left(A^2\frac{\partial U}{\partial\phi}\delta Q + \dot\phi\ (\delta Q)\dot{}\right) +4\dot{\Psi}\dot\phi-2\Psi A^2\frac{\partial U}{\partial\phi},\\
(\delta Q)\ddot{} &+& 3H(\delta Q)\dot{}+\left(\frac{k^2}{a^2}+\frac{\partial^2V}{\partial Q^2}-2\beta^2A^2\dot{\phi}^2+16 A^4\beta^2 U\right)\delta Q \nonumber\\ &=& -2\Psi\left(\frac{\partial V}{\partial Q}+4A^4 U\beta\right)+4\dot\Psi\dot Q + 2A^2\beta\dot\phi (\delta\phi)\dot{} - 4A^4\beta\frac{\partial U}{\partial\phi}\delta\phi.
\end{eqnarray}

\begin{figure}[htp!]
\centering
\includegraphics[width=14cm]{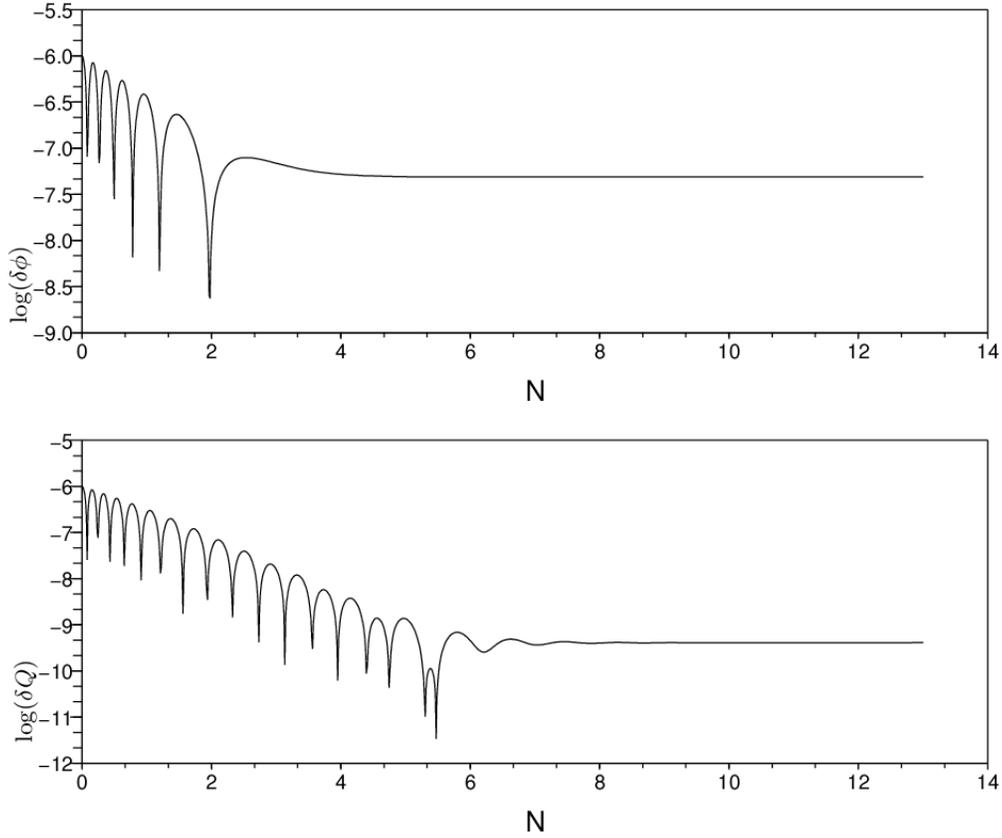}
\caption{The evolution of the field perturbations as a function of $N=\ln a$. The upper panel shows the perturbations of the inflaton and the lower panel shows the perturbations of the $Q$-field. Here $\beta = 0.5$, $\lambda=10$, $m=7\times 10^{-7}$ and $k=0.001$. As one can see, the perturbations in $Q$ drop to a value several orders of magnitude below that of the perturbations in $\phi$, due to the heaviness of the $Q$-field relative to $\phi$.}
\label{fig:perts}
\end{figure}

Integrating these equations numerically shows the perturbations of the $Q$-field are indeed suppressed. 
A typical plot of the evolution of $\delta\phi$ and $\delta Q$ is shown in Fig.~\ref{fig:perts}. 
As already said above, the difference in the behaviour of the two quantities is due to the large effective mass of the $Q$-field. 
Thus, the inflaton perturbations dominate and we can ignore the perturbations 
of the quintessence field. We have checked numerically that in the regime we are interested in, 
the perturbations in $Q$ are always suppressed relative to the perturbations in $\phi$. There is an intermediate regime, 
in which the quintessence mass is smaller but of order $H$ and contributes to the cosmological expansion by a reasonable amount. 
In this case, the quintessence field will not sit in the minimum of the effective potential, but is 
attracted to it and its perturbations cannot be ignored. However, in this paper we do not deal with this case. 

With this in mind, the calculations proceed in the standard way, taking into account the modifications 
of the background evolution, as discussed above. The power spectrum of curvature perturbations is
\begin{equation}\label{eq:curvspec}
{\cal P}_{\cal R}(k) = \left(\frac{H}{\dot{\phi}}\right)^2\left(\frac{H}{2\pi}\right)^2 \approx \frac{1}{24\pi^2 M_{Pl}^4}\frac{U A^6}{\epsilon}\left(1+\frac{4\beta}{\lambda}\right).
\end{equation}

The spectral index is found by calculating the derivative of this quantity  with respect to ${\rm ln}(k)$, where 
$k$ is the wavenumber. Using the slow-roll condition one finds
\begin{equation}
\frac{d}{d\ln k} = -M_{Pl}^2\frac{U'}{U A^2}\left(1+\frac{4\beta}{\lambda}\right)^{-1}\frac{d}{d\phi}.
\end{equation}
The derivatives of the slow-roll parameters are: 
\begin{eqnarray}
\frac{d\epsilon}{d\ln k} &=& -2\eta\epsilon +4\epsilon^2\left(1+\frac{3\beta}{\lambda}\right), \\
\frac{d\eta}{d\ln k} &=& -\xi^2 +2\eta\epsilon\left(1+\frac{2\beta}{\lambda}\right),
\end{eqnarray}
where
\begin{equation}
\xi^2\equiv \frac{M_{Pl}^4}{A^4}\frac{U'''U'}{U^2}\left(1+\frac{4\beta}{\lambda}\right)^{-2}.
\end{equation}
So we find the spectral index to be
\begin{equation}
n_s-1 = 2\eta-6\epsilon-\frac{8\beta}{\lambda}\epsilon.
\end{equation}
The running of the spectral index is found to be
\begin{equation}
\frac{d n_s}{d\ln k}=-2\xi^2+16\eta\epsilon-24\epsilon^2+\left(\frac{24\beta}{\lambda}\eta\epsilon
-\frac{104\beta}{\lambda}\epsilon^2-\frac{96\beta^2}{\lambda^2}\epsilon^2\right).
\end{equation}
For the tensor perturbations, we have 
\begin{equation}
{\cal P}_{\rm grav} = \frac{2}{M_{\rm Pl}^2}\left(\frac{H}{2\pi}\right)^2. 
\end{equation}
Writing ${\cal P}_{\rm grav}\propto k^{n_g}$, we can show that 
\begin{equation}
n_g = -2\epsilon,
\end{equation} 
whereas the tensor to scalar ratio is, using the slow-roll condition (\ref{eq:slowroll1}), 
\begin{equation}\label{eq:tsratio}
r \equiv \frac{{\cal P}_{\rm grav}}{{\cal P}_{\cal R}} = \frac{4 \epsilon}{A^2} = -\frac{2 n_g}{A^2}
=- 2 n_g \left(\frac{4\beta U}{\lambda M_{\rm Pl}^4}\right)^{2\beta/(4\beta + \lambda)}.
\end{equation}
Note that while the prediction for the scalar-tensor ratio $r$ is modified from the standard case, 
the expression for the tensor spectral index, $n_g$, remains the same. 

\subsection{Consequences} \label{sec:cons}
Having derived the relevant equations describing slow-roll inflation with a coupled 
dark energy scalar field, we will now discuss the consequences and predictions. 
The first difference from the standard case is that the expressions for the slow-roll parameters have changed. 
The origin of the modifications are two-fold: firstly, the slow-roll condition for the 
inflaton field is modified and secondly, the expansion rate is enhanced by a factor 
$A^4(1+4\beta/\lambda)$. As a result, both slow-roll parameters contain an additional factor $A^{-2}$ 
when compared to the standard expression, see eqns. (\ref{eq:epsilon}) and (\ref{eq:eta}). Additionally, 
they are modified by a factor $(1+4\beta/\lambda)^{-2}$ (for $\epsilon$) and 
$(1+4\beta/\lambda)^{-1}$ (in the case of $\eta$). Slow-roll inflation ends 
when $\rm{max} \{\epsilon ,\eta\}= 1$. The presence of the factor of $A^{-2}$ means that 
the end of inflation is delayed in this model, in the sense that smaller values of $\phi$ are reached in the 
slow-roll phase. An immediate consequence of this is that the oscillations of the inflaton around its minimum -- 
responsible for the production of particles in the reheating phase -- will have a smaller amplitude than in the 
standard case. 

Another important difference is in the expressions for the cosmological perturbations. Apart from the 
spectral index of the gravitational wave power spectrum, all expressions for the perturbations 
have changed: the amplitude of scalar perturbations is different and the expressions for the 
spectral index $n_s$ and its running include additional factors which depend on the ratio $\beta/\lambda$. 
However, because the actual values of $\epsilon$ and $\eta$ during the last 60 efolds change from 
their values in the standard chaotic inflationary scenario, it is 
not obvious whether $n_s$ will be bigger or smaller than in the standard case. 

\begin{figure}[htp!]
\centering
\subfigure[$\log(m/M_{\rm Pl})$ as a function of $\lambda$ and $\beta$]
{
\includegraphics[width=6.5cm]{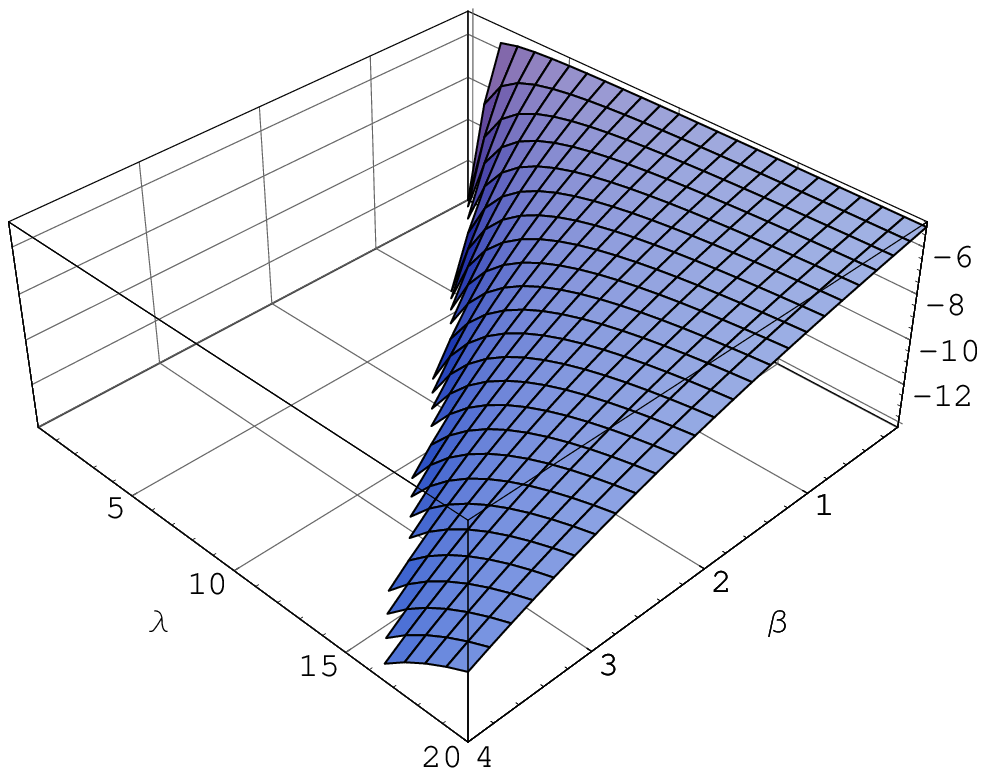}
\label{fig:mvalues}
}
\hspace{0.5cm}
\subfigure[$n_s$ as a function of $\lambda$ and $\beta$]
{
\includegraphics[width=6.5cm]{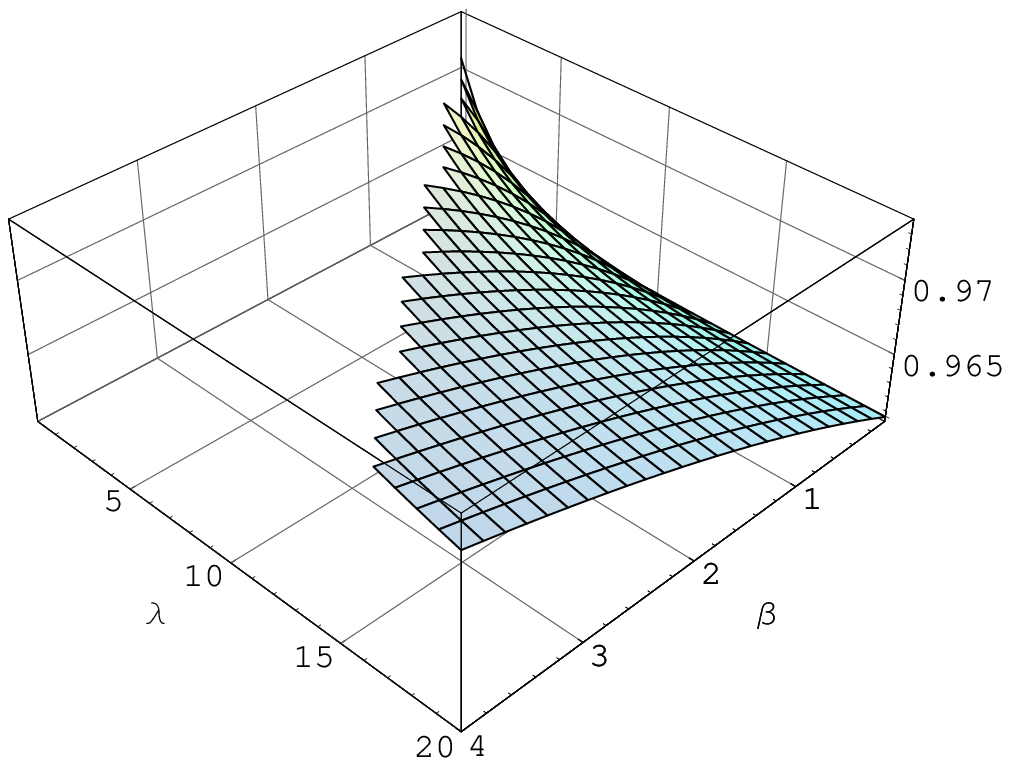}
\label{fig:nvalues}
}
\caption{Predictions for a chaotic inflationary potential: $U=m^2\phi^2 / 2$. 
The values of the inflaton mass (as obtained by normalizing the power spectrum to 
COBE) and spectral index vary with the parameters $\lambda$ and $\beta$. 
In these plots, only the data for $\lambda > 4\beta$ (i.e. the region in which the inflaton is the dominant 
component during inflation) are plotted.}
\end{figure}

\begin{figure}[htp!]
\centering
\includegraphics[width=14cm]{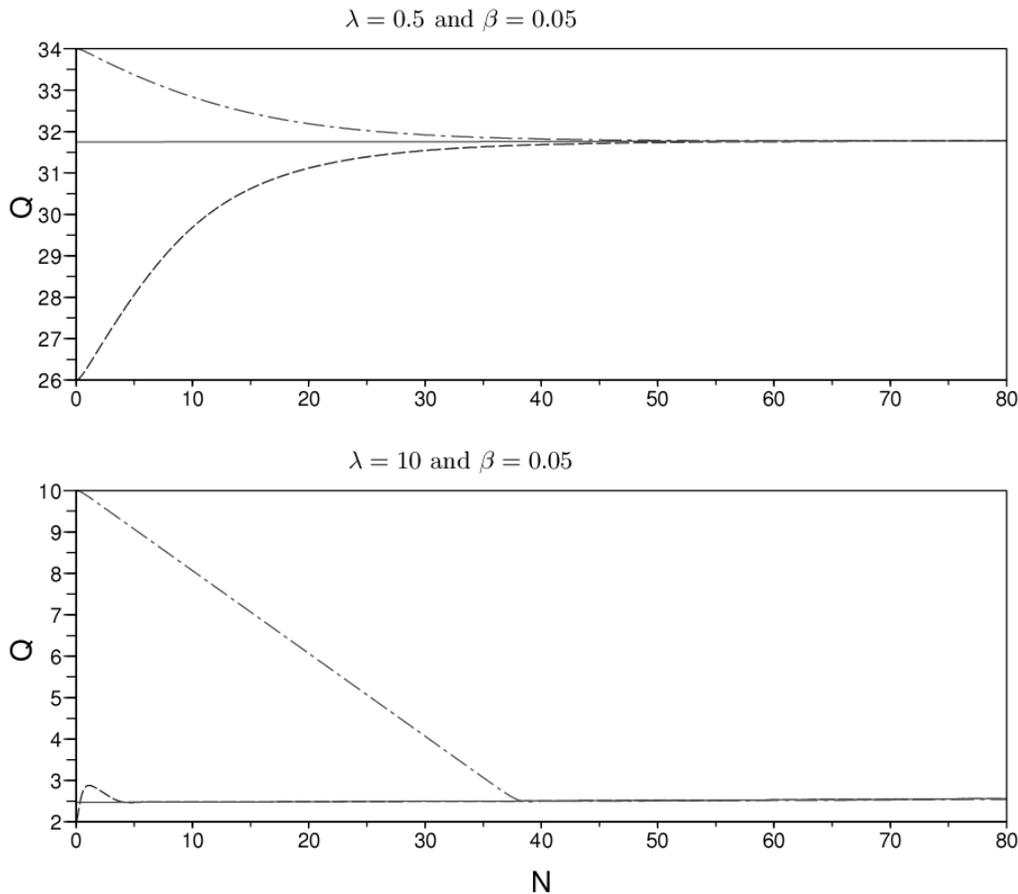}
\caption{These plots show the $Q$-field as a function $N=\ln a$ (dotted lines). Even if the initial value of the field changes, the field is drawn toward its minimum (solid line).}
\label{fig:min}
\end{figure}

To illustrate the effect of the dark energy field, let us consider as an example the case of a massive 
inflaton field with potential energy given by eq. (\ref{eq:quad}). The theory has three free parameters, 
namely $m$, $\beta$ and $\lambda$. We can 
therefore use eq.(\ref{eq:curvspec}) and the COBE normalization to fix $m$ as a function of $\beta$ 
and $\lambda$.  The results are 
shown in Fig. \ref{fig:mvalues}. At the same time, we have to make sure that $\beta$ is big enough to ensure that the 
$Q$--field sits in the minimum of the effective potential. We have checked numerically that for 
$\beta \geq 0.05$, the field sits indeed in the minimum of the effective potential (see fig. \ref{fig:min}). As one can see, the mass of the inflaton field is smaller than in the standard 
case ($m\approx 10^{-5}M_{\rm Pl}$). This is because of the factor $A^6$ in the expression for the 
power spectrum and, since $A\geq 1$, a smaller mass $m$ is sufficient to obtain the correct amplitude 
for the power spectrum. The predictions for the spectral index $n_s$ are shown in Fig \ref{fig:nvalues}. 
As one can see, the spectral index $n_s$ is larger than in the standard case without couplings. 
The current constraint from WMAP alone on the spectral index is $n_s=0.963^{+0.014}_{-0.015}$, 
assuming a $\Lambda$CDM cosmology \cite{Dunkley:2008ie}. Note that in order to compare the theory to data,
one would have to study the evolution of cosmological perturbations 
in the presence of dark energy couplings in the subsequent epoch and make assumptions about the 
couplings to different matter species. The coupling $\beta$ we have discussed so far is the coupling of dark energy to 
the inflaton field and is {\it a priori} not the same as the coupling to dark matter or neutrinos. If 
dark energy is subsequently only coupled to dark matter, then current constraints give $\lambda \leq 0.95$ and 
$|\beta| \leq 0.055$ at 95\% confidence level \cite{Bean:2008ac}. Assuming that the dark energy coupling to the inflaton 
field is the same as to dark matter, this implies that the predictions of the dark energy coupling to the 
primordial perturbations are only slightly modified from the standard chaotic inflation predictions. On the 
other hand, in the case of theories with a mass-growing component \cite{Amendola:2007yx}, which requires larger values for $\lambda$ and $\beta$, 
modifications to the primordial power spectrum are predicted. In any case, when comparing the theory to data, 
assumptions about the subsequent composition 
of the universe and new interactions between dark energy and other matter forms have to be made. The 
results presented in this section will be useful when constructing a theory of inflation, dark matter and 
dark energy based on particle physics. 

\section{Reheating and Preheating in the Presence of Coupled Quintessence} \label{sec:3}
In our scenario, when the slow-roll conditions are violated, the value of $\phi$ is smaller than in the standard case and thus the oscillations of  the field around its minimum are of smaller amplitude.  This can affect the 
processes of reheating and preheating. In the 
following we will discuss in detail the effect that the presence of coupled quintessence has on these important epochs.

\subsection{Reheating} \label{reheating}

Any successful model of reheating relies heavily on the oscillatory behaviour of the inflaton. 
The oscillations are damped due to the expansion of the Universe by a friction term proportional to $H$ and 
a term describing the decay of the inflaton field into radiation. Following the standard treatment of reheating (cf. \cite{KolbTurner}), the equation of motion for the inflaton field 
reads
\begin{equation}
\ddot{\phi} + (3H+\Gamma_{\rm r})\dot{\phi}+A^2m^2\phi =  -2 \beta \dot Q \dot\phi,
\end{equation}
where $\Gamma_{\rm r}$ is the decay rate of the process.
In the standard single-field case the value of the Hubble parameter is proportional to the square root of the total energy density, 
which consists of inflaton energy density and radiation energy density. In the present case, however, there is also the 
energy density of the $Q$-field, which contributes to the Hubble friction. Just at the end of inflation, for example, the ratio 
of the energy densities of quintessence and the inflaton field is roughly $4\beta/\lambda$, from equation (\ref{minconst}). This 
ratio can be potentially quite big (up to $10^{-1}$, or so), so the $Q$-field contributes significantly to the 
Hubble damping. This could cause heavy damping of the inflaton oscillations that would reduce the efficiency 
of the reheating process.

\begin{figure}[htp!]
\centering
\includegraphics[width=14cm]{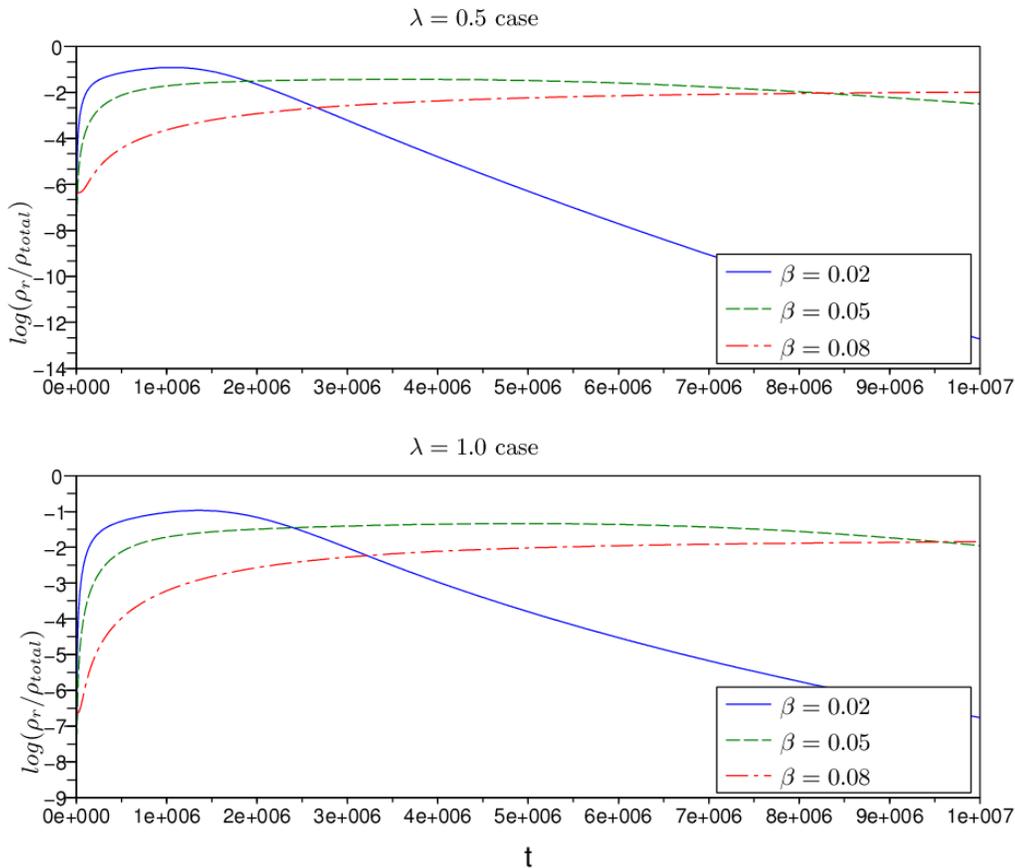}
\caption{The log of the ratio of the radiation energy density to the total energy density for different values of $\lambda$ and $\beta$. It is not possible to get radiation dominance using constraints on the parameters from \cite{Bean:2008ac}. We have restricted ourselves to the case where the coupling, $\beta$, is large enough make $Q$ stay in the minimum of its effective potential.}
\label{fig:lam_bet1}
\end{figure}

In the model of elementary reheating, the equation of motion for the energy density of  radiation, $\rho_{r}$, is
\begin{equation} \label{eq:rad}
\dot{\rho}_{\rm r}+4H\rho_{\rm r}=\Gamma_{\rm r}\left(\rho_\phi + p_\phi \right), 
\end{equation}
In the standard scenario without quintessence, 
the energy density of the radiation produced from the 
decay of the inflaton quickly increases to a maximum value, then decreases as the decay products are diluted 
by the expansion of the universe. This continues until $H\approx\Gamma_{\rm r}$, when the inflaton decays away rapidly and 
the radiation-dominated era begins with temperature, $T_{\rm RH}$. In our  model, we want to avoid the dominance of 
the $Q$-field, and it is therefore necessary either to produce enough radiation so that $[\rho_{rad}]_{MAX}$ is greater 
than  $\rho_Q$ or to dissipate the energy of the $Q$-field quickly, so that it becomes smaller than $\rho_{r}$. 
$[\rho_{r}]_{MAX}$ depends on the  decay rate, $\Gamma_{\rm r}$ and therefore on   the details of the decay processes. For example, if the field decays to two light scalar particles with coupling $gA$, the highest decay rate is $\Gamma_{\rm r}\sim mA$ \cite{Mukhanov}. This means that the amount of radiation produced is limited by the (effective) inflaton mass, the value of which, as was seen in Section \ref{sec:2}, must be chosen to be consistent with the COBE normalisation for particular values of $\lambda$ and $\beta$. Figure \ref{fig:lam_bet1} shows that if one uses constraints on the Quintessence potential from \cite{Bean:2008ac}, it is not possible to get successful reheating in this model as $\rho_{\rm r}$ is always less than $\rho_Q$.

\begin{figure}[htp!]
\centering
\includegraphics[width=14cm]{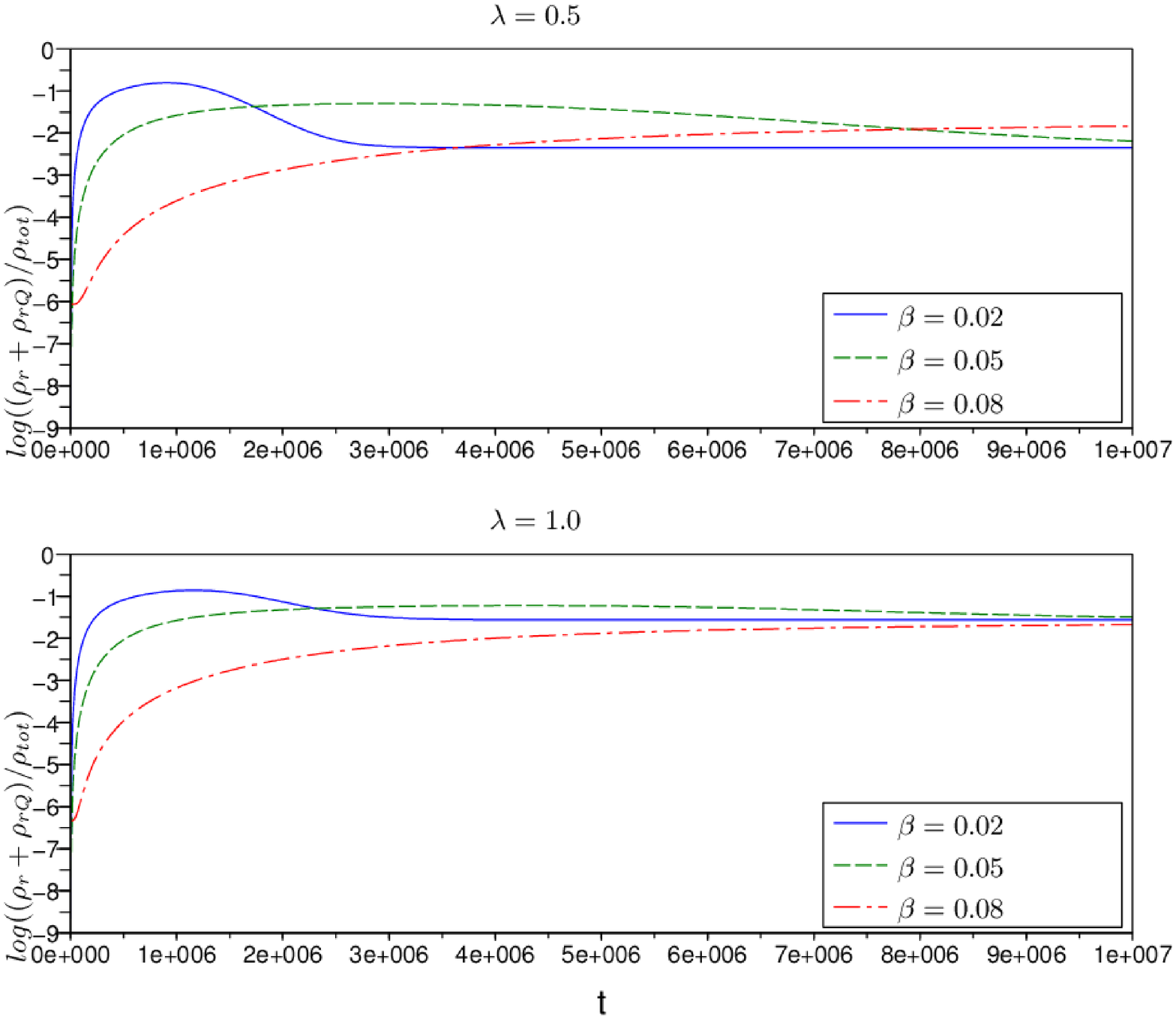}
\caption{In this simulation, the model was modified to allow the $Q$-field to decay to radiation in a manner analogous to $\phi$ by adding a friction term $\Gamma_{\rm Q} \sim m_{\rm eff}$ to eqn.~(\ref{eq:Qeq}). The radiation produced in this manner contributes a fraction of the total energy density. These plots shows the log of the fraction of the energy density accounted for by the total amount of radiation (i.e. including also that from the decay of the inflaton).}
\label{fig:Qdecay}
\end{figure}

We are thus forced to  reduce the energy density of the $Q$-field. One method would be to let $Q$ decay into radiation in a similar way to $\phi$, with a decay rate proportional to the effective mass of the field. This would introduce a term $\rho_{\rm rQ}$, representing the energy density of the radiation produced in this manner. In this case the equations for $Q$ and $\rho_{\rm rQ}$ are:
\begin{eqnarray}
\ddot Q + (3H+\Gamma_{\rm Q}) \dot Q + \frac{\partial V}{\partial Q} = A^2 \beta (\dot\phi^2 - 4A^2 U(\phi)), \\
\dot{\rho}_{\rm rQ}+4H\rho_{\rm rQ}=\Gamma_{\rm Q}\left(\rho_Q + p_Q \right)
\end{eqnarray}
 However, numerical calculations (see fig. \ref{fig:Qdecay}) show that the radiation produced is still not sufficient. This is partly due to the small magnitude of the term driving the production of radiation: $\Gamma_{Q}\left(\rho_Q + p_Q \right) = \Gamma_{Q}\dot{Q}^2$. ($\dot{Q}$ is small as the effective potential is very flat at this stage.) 

\begin{figure}[htp!]
\centering
\includegraphics[width=14cm]{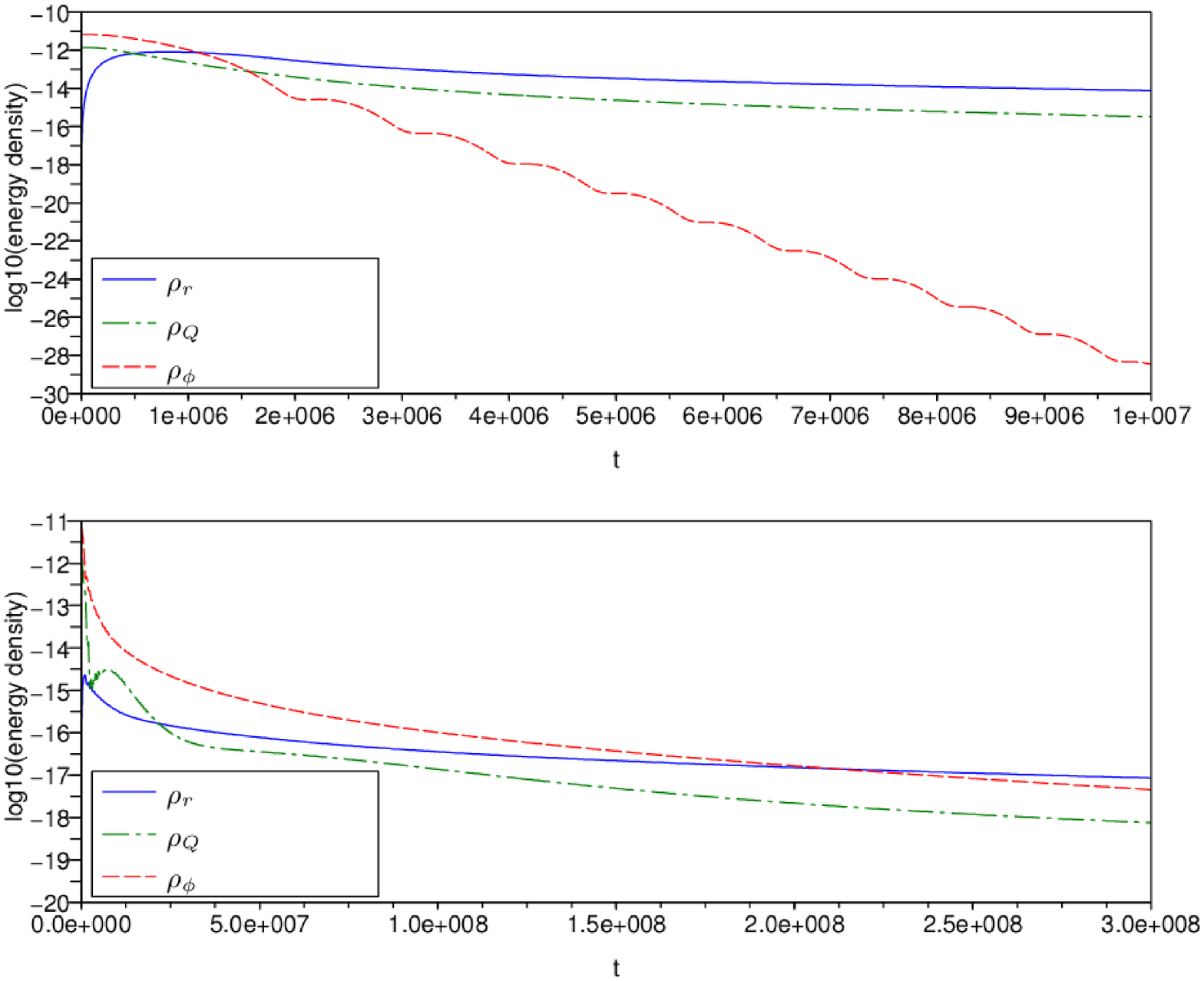}
\caption{These plots show the evolution of $\rho_\phi$, $\rho_Q$ and $\rho_{\rm r}$ (on a log scale) for large $\lambda$. Here, $\lambda=10$, $\beta=0.5$ and $m=7\times 10^{-7}$. In the upper panel, the decay rate is large ($\Gamma_{r} =0.9mA$) so radiation domination occurs very quickly. In the lower panel, although the decay rate is much smaller ($\Gamma_{r} =0.001mA$), the radiation can still dominate over the $Q$-field but this process is of longer duration. The reheating temperature would be smaller in this latter case.}
\label{fig:biglam}
\end{figure}

A favourable alternative is to vary the parameter $\lambda$ that controls the steepness of the potential of the $Q$-field. A larger $\lambda$ means that the field rolls more quickly to larger values. This has a twofold effect: $[\rho_{rad}]_{MAX}$ is increased (as the damping is reduced, so the source term in eqn. (\ref{eq:rad}) does not
 decrease as quickly as in the previous case) and $\rho_Q$ is decreased, so $[\rho_{rad}]_{MAX}$ does not have to be very  large for radiation to dominate the universe. It can be seen in fig. \ref{fig:biglam} that using a large value of $\lambda$ does indeed allow radiation to dominate the universe. It is interesting to note that this is true even if $\Gamma_{\rm r}$ is much less than the maximum possible decay rate. In this case, as can be seen in the lower panel of fig. \ref{fig:biglam}, the total amount of radiation produced is dramatically reduced, leading to a smaller reheating temperature than in the case of large $\Gamma_{\rm r}$. The same result is true in the standard single field case, where $T_{RH}\approx0.55g_{*}^{-1/4}(\sqrt{8\pi}M_{Pl}\Gamma_\phi)^{1/2}$ \cite{KolbTurner}. The difference is that in our model if the decay rate is {\it too} small, one cannot satisfy the condition $\rho_{\rm r} > \rho_Q$.

It is clear that if our model is to be consistent, the parameters of the quintessence field must be constrained by this theoretical consideration and $\lambda$ must be large.
 It has been shown \cite{Amendola:2007yx} that it is possible to achieve a realistic cosmology with large $\lambda$ in the case with a growing matter component. As Figure \ref{fig:biglam} shows, it is possible for radiation to dominate the universe after the decay of the inflaton using values that satisfy this lower bound.  

\subsection{Preheating} \label{sec:preheating}
We have seen that reheating is considerably affected by the presence of a coupled quintessence field. 
In particular, 
the process is less efficient than in the standard case, because the Hubble damping is much larger. In 
the following, we discuss the process of preheating, in which the inflaton energy is converted into other particles 
in a very efficient way. We consider the possibility that the energy density of these decay products could 
counterbalance the effect of the $Q$-dominance and lead to successful reheating. 

Following the standard description of preheating, as discussed in \cite{Kofman:1997yn,Kofman:1994rk,Lachapelle:2008sy}, we assume that the inflaton 
field decays to a light scalar field $\chi$ of mass $m_\chi$. To describe the interaction between $\phi$ and $\chi$, we 
will add the following term in the Lagrangian:
\begin{equation}\label{eq:coupling}
\Delta {\cal L}_{\rm int} = -g A \phi\chi^2. 
\end{equation}
The conformal factor $A$ arises because our analysis is performed in the Einstein frame. This term describes a three-legged 
interaction involving one inflaton particle and two $\chi$ particles. Another common choice of interaction is of the form
$-\textstyle\frac{1}{2} g^2 \phi^2 \chi^2$. As similar behaviour arises in both cases and we are interested in the more 
general consequences of the presence of the $Q$-field on the mechanism of reheating, we will concentrate our attention on 
the interaction in eqn.~(\ref{eq:coupling}). We assume that the effective coupling constant $gA$ is less than the frequency 
of the inflaton oscillations $mA$, so that the interaction is not modified too much by quantum corrections. 
Thus, we have the limit $g<m$. The vacuum expectation value of the $\chi$-field is zero, so the Friedmann 
equation and the classical equation of motion for the inflaton will be unaffected. Expanding $\chi$ around zero, we find 
the following equation for the perturbations of $\chi$, which are interpreted as particles after quantization:
 
\begin{equation}\label{eq:fullchi}
\ddot{\chi_k}+3H\dot{\chi_k}+\left(\frac{k^2}{a^2}+m_\chi^2+2gA\phi\right)\chi_k = 0.
\end{equation}
To study the resonance, we neglect the expansion of space and introduce a sinusoidal ansatz with which to 
describe the oscillations of the inflaton, $\phi = \Phi \sin(mAt)$. In this case, eqn.~(\ref{eq:fullchi}) 
can be written in the form of a Mathieu equation, i.e. 
\begin{equation} \label{eq:Mathieu}
\chi_k''+[A_k-2q\cos(2z)]\chi_k = 0,
\end{equation}
with
\begin{eqnarray}
A_k &\equiv& \frac{4(k^2+m_\chi^2)}{m^2A^2}, \label{eq:Ak} \\
q &\equiv& \frac{4g\Phi}{m^2A}, \label{eq:q}\\
z &\equiv& mAt/2,
\end{eqnarray}
where prime indicates a derivative with respect to $z$. These are the standard equations with the addition of 
the factors $A$ in the expressions for $A_k$, $q$ and $z$ (cf. \cite{Mukhanov} or \cite{Kofman:1997yn}). As we have also seen in Section \ref{sec:2}, the 
inflaton mass is smaller than in the standard theory once the theory is normalized to COBE and 
the amplitude $\Phi$ of the inflaton field is smaller. We note that for a {\it given} value of $g$, the 
parameter $q$, which determines the behaviour of the solutions to the equation above, 
can be considerably bigger than in the standard case. 

A similar equation can be derived for the $Q$-particles produced, since the $Q$-field couples to the inflaton field as well. 
We obtain ($Q_k$ is the Fourier component of the perturbation of the $Q$-field around its VEV)
\begin{equation}
Q_k''+[A_k^{(Q)}-2q^{(Q)}\cos(2z_Q)]Q_k = 0,
\end{equation}
with
\begin{eqnarray}
A_k^{(Q)} &\equiv& \frac{(k^2+m_Q^2)}{m^2A^2}+\frac{6}{5}q^{(Q)},\\
q^{(Q)} &\equiv& \frac{5}{2} \Phi^2 \beta^2 A^2,\\
z_Q &=& mAt.
\end{eqnarray}
Since $m_Q$ is much bigger than $mA$, we see that $A_k^{(Q)}$ is dominated by the first term and therefore $A_k^{(Q)}\gg q^{(Q)}$. 
This means that the periodic term in the equation for $Q_k$ does not play an important role and there is no resonance production of 
$Q$ particles. Therefore we concentrate on the production of the light $\chi$ particles in the following. 

Let us consider first the case of narrow resonance, in which $q \ll 1$. Solutions to eq.~(\ref{eq:Mathieu}) 
falling within particular resonance bands are exponentially unstable and take the form $\chi_k \sim \exp(\mu_k z)$. 
The most important band is the first one, for which the resonance reaches its maximum at $\mu_k \approx q/2$. 
The number density of particles with momentum $k$ is
\begin{equation}\label{eq:numbers}
n_k=\frac{\omega_k}{2}\left(\frac{|\dot{\chi}|^2}{\omega_k^2}+|\chi|^2\right)-\frac{1}{2},
\end{equation}
where $\omega_k^2 = A_k-2q\cos(2z)$. From this, one can see that,while narrow resonance continues, 
the number of particles with $k \approx mA/2$ increases exponentially, with $n_k \sim \exp(qz)$ 
(as $\chi$ is, by design, much lighter than the inflaton and the resonance occurs at $A_k \sim 1$).

Narrow resonance will only be an important decay mechanism as  long as it is more efficient than 
the perturbative decay, which corresponds to the elementary theory of reheating discussed in the last 
section. In narrow resonance, the number of particles in the resonance band $k$ increases 
exponentially as $n_k \sim \exp(qz) \sim \exp(qmAt/2)$ so the decay rate can be approximated by 
$qmA$. In the regime where perturbative decay itself is efficient (i.e. $\Gamma_\chi > H$), 
we find that the condition for narrow resonance to be the leading effect is
\begin{equation}
qmA > \Gamma_\chi, \label{eq:cond1}
\end{equation}
where $\Gamma_\chi = g^2 A/(8\pi m)$ is the decay rate as obtained by standard quantum field theory methods.  
The second phenomenon that affects the timescale of narrow resonance is the redshift of momenta out of the resonance bands. The exponential nature of the resonance means that the rate of production of $\chi$ particles depends on the number of particles present already. If the modes do not spend enough time within the resonance band, $n_k$ will be small and the process will be inefficient. 

The width of the resonance band at $k=mA/2a$ is $kq$. If we assume that the resonance is most efficient in the middle of this band we have $\Delta k = mAq/2$ . We wish to find the time, $\Delta t$, that the mode spends in the region $\Delta k$. Using $\Delta k = |\textstyle\frac{dk}{dt}| \Delta t = kH\Delta t$, this is $\Delta t = qH^{-1}$. During this interval, the number of particles increases as $\exp (qmA\Delta t /2) = \exp (q^2 mA/2H)$. So the second condition required for decay by narrow resonance to be efficient is
\begin{equation}
q^2 mA > H. \label{eq:cond2}
\end{equation}
We can write the inequalities in terms of the inflaton amplitude, $\Phi$.
\begin{eqnarray} \label{eq:cond3}
\Phi &>& \frac{gA}{32\pi}, \\
\Phi &>& \frac{m\sqrt{mAH}}{4g}.\label{eq:cond4}
\end{eqnarray}

 A small value of $m$ means that parametric resonance can continue unhindered for long enough to produce large amounts of $\chi$ particles, which subsequently decay to radiation. 

\begin{figure}[htp!]
\centering
\includegraphics[width=14cm]{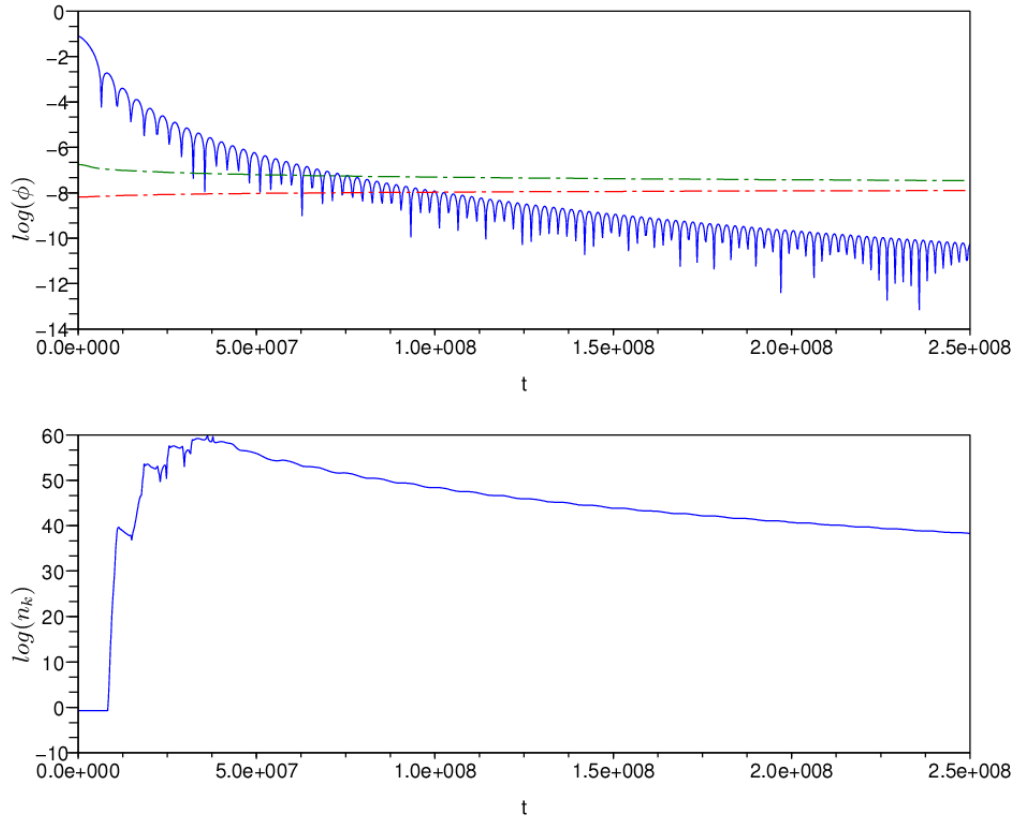}
\caption{The oscillations of the $\phi$ field are still capable of driving parametric resonance, despite their small amplitude. In this plot, $\beta=0.1$ and $\lambda=1$. $g$ is set to $0.9m$, where $m$ is the inflaton mass, which is chosen to be $4\times 10^{-8}$ to match the COBE normalisation. In upper plot we show the behaviour of $\phi$, whereas the lower plot shows the number density $n_k$ of the $\chi$ particles. The dotted lines on the upper plot show the bounds given in eqns. (\ref{eq:cond3}) and (\ref{eq:cond4}), below which narrow resonance is not possible.}
\label{fig:broad1}
\end{figure}

If, instead, the parameter $q$ is greater than one, we are in the regime of broad resonance. Here, the mass 
term in eqn. (\ref{eq:Mathieu}) is dominated by the sinusoidal behaviour of the inflaton. When this term is zero, 
the number of particles given by eq.~(\ref{eq:numbers}) is not conserved and one observes resonance in a broad 
range of modes. This can be seen in figure \ref{fig:broad1}. The small value of $m$ required to match the COBE normalisation means that the parameter $q$ is very large in this case, leading to a large increase in the number of $\chi$ particles, despite the small amplitude of the oscillations.

The theory governing the energy density of the radiation produced by this method and the backreaction of the particles on the evolution of the inflaton is dependent on the model with which one is working and the free parameters therein. We can assume, at the very least, that the energy density of the products of parametric resonance will not exceed that of the inflaton prior to decay. As, at the start of the reheating era, $\rho_Q \approx \left(\frac{4\beta}{\lambda}\right)\rho_\phi$, Hubble damping of the inflaton can reduce the energy density of the inflaton field to a value less than that of the $Q$-field before the field can complete many oscillations about zero. This rapid damping is exhibited in figure \ref{fig:broad1}. Because any model of parametric resonance assumes the inflaton is oscillating, by the time a significant quantity of $\chi$-particles are produced, the inflaton is incapable of producing enough radiation for it to dominate over the $Q$-field. Therefore, even if we employ parametric resonance, we still have to use a large value of $\lambda$ to reheat the universe in our model of coupled quintessence. 

\section{Conclusions} \label{sec:conclusions}

We have described a model of coupled quintessence in which the scalar field responsible for dark energy is coupled to the inflaton field. To be specific, we have used the exponential potential in eqn. (\ref{eq:pot}) and assumed that the coupling between the fields is of the form $A(Q)=\exp(\beta Q/M_{Pl})$ with $\beta$ positive, so that the effective potential has a minimum. The large effective mass of the $Q$-field drives it into this minimum for a large range of $\lambda$ and $\beta$ values, and in Section \ref{sec:21} we used this to evaluate the slow-roll parameters. We found that the expressions for $\eta$ and $\epsilon$ are modified by factors of $A^{-2}(1+4\beta/\lambda)^{-1}$ and $A^{-2}(1+4\beta/\lambda)^{-2}$ respectively, meaning that the end of inflation is delayed in this model, so that the inflaton field has a smaller field amplitude during its oscillations. 

We further discussed the modification to the cosmological perturbations. The equations for the spectral index and its running (in terms of the new slow-roll parameters) vary with the ratio $\beta/\lambda$. Normalising the power spectrum to COBE, we found that the mass of the inflaton field is required to be smaller than in the standard case, whilst the effect of the presence of a dark energy coupling is to increase the value of $n_s$, for a given $\lambda$. The parameters $\beta$ and $\lambda$ can be constrained by observations if one assumes that the coupling between the inflaton and dark energy has the same equal magnitude as the coupling between dark matter (or neutrinos) and dark energy. 

We found that the presence of the $Q$ field during reheating leads to an unnaturally large amount of Hubble damping in the equation for the inflaton oscillations during reheating. We showed that if reheating is to be successful, $\lambda$ must be large. The dark energy field rolls faster down its potential, which in turn reduces the Hubble damping and allows the radiation produced during reheating to become the dominant component of the universe. We also considered how the mechanism of parametric resonance during preheating might work in this model. We found that parametric resonance could still produce large amounts of radiation as the smaller value of $m$ required to match the COBE normalisation decreases the lower bound on the amplitude of the inflaton oscillations required for resonance. However, if $\lambda$ is small, the energy density of the inflaton quickly becomes less than that of the $Q$-field so a large value of $\lambda$ is still required to reheat the universe in our model.
The fact that a large $\lambda$ is required for successful reheating has interesting consequences. For example, it is not compatible with the condition $\lambda<0.95$ stated in \cite{Bean:2008ac}. Such models would be only consistent if the quintessence field does not couple to the inflaton field. 
Alternatively, the dark energy field could couple to a subdominant species, such as neutrinos, as e.g. in \cite{Amendola:2007yx}. 

\vspace{1cm}

\noindent{\bf Acknowledgements:} JW is supported by EPSRC, CvdB and LHMH are supported by STFC. CvdB thanks the 
Institut de Physique, Saclay and Christof Wetterich and the Institut f\"ur Theoretische Physik of the University 
of Heidelberg for hospitality while parts of this work have been completed. 

\bibliographystyle{unsrt}
\bibliography{refs}

\end{document}